\documentclass[journal=ancac3,manuscript=article]{achemso}

\usepackage[version=3]{mhchem} 
\usepackage{siunitx}

\author{Lukas Worch}
\affiliation[1]{Department of Materials, London Centre for Nanotechnology, Royal School of Mines, Imperial College London, United Kingdom}
\author{Kavin Arunasalam}
\affiliation[1]{School of Chemistry, CRANN and AMBER Research Centres, Trinity College Dublin, Dublin 2, Ireland}
\author{Neil Mulcahy}
\author{Syeda Ramin Jannat}
\author{James Douglas}
\affiliation[1]{Department of Materials, London Centre for Nanotechnology, Royal School of Mines, Imperial College London, United Kingdom}
\author{Baptiste Gault}
\affiliation[1]{Department of Materials, London Centre for Nanotechnology, Royal School of Mines, Imperial College London, United Kingdom}
\alsoaffiliation[3]{Max Planck Institute for Sustainable Materials, Max-Planck-Str. 1, 40237 D\"usseldorf, Germany}
\alsoaffiliation[4] {now at Univ Rouen Normandie, CNRS, INSA Rouen Normandie, Groupe de Physique des Matériaux, UMR 6634, F-76000 Rouen, France}
\author{Valeria Nicolosi}
\affiliation[2]{School of Chemistry, CRANN and AMBER Research Centres, Trinity College Dublin, Dublin 2, Ireland}
\author{Michele Shelly Conroy}
\affiliation[1]{Department of Materials, London Centre for Nanotechnology, Royal School of Mines, Imperial College London, United Kingdom}
\email{mconroy@imperial.ac.uk}

\title[An \textsf{achemso} demo]
  {Probing the Liquid Solid Interfaces of 2D SnSe MXene Battery Anodes at the Nanoscale}

\abbreviations{}

\begin{document}



\begin{abstract}

Understanding degradation processes in lithium ion batteries is essential for improving long term performance and advancing sustainable energy technologies. Tin selenide (SnSe) has emerged as a promising anode material due to the high theoretical capacity of tin. Unlike conventional intercalation based electrodes, SnSe undergoes conversion and alloying reactions with lithium to form Li$_{4.4}$Sn, Sn, and Li$_2$Se, enabling high lithium storage but inducing large volume changes that cause mechanical instability and capacity fading. Embedding SnSe nanoparticles within a Ti$_3$C$_2$T$_x$ MXene framework offers a strategy to mitigate these effects by enhancing conductivity and structural resilience.

Here, cryogenic focused ion beam (cryo FIB) slice and view revealed progressive material redistribution and morphological transformation during cycling, underscoring the need for site specific chemical analysis. Cryogenic atom probe tomography (cryo APT) of selected regions provided high spatial and chemical resolution while preserving beam sensitive phases, uncovering nanoscale degradation mechanisms including phase transformations, partial dissolution of active material, and, importantly, the first direct evidence of copper corrosion and copper ion migration from the current collector into the electrode. The observation of copper redistribution demonstrates that current collector degradation contributes directly to chemical contamination and capacity fading in composite electrodes. Together, cryo FIB and cryo APT provide a powerful workflow for elucidating electrode degradation in reactive, beam sensitive systems, offering critical insights for designing more durable and stable next generation battery materials.

\end{abstract}

\section{Introduction}

Progress in lithium ion battery (LIB) performance, including specific capacity, power, and cycling stability, fundamentally depends on the design of advanced electrode materials and the stability of the current collector. \cite{janek2023challenges, xiao2023laboratory, wang2022toward, acebedo2023current, nasajpour-esfahani_comprehensive_2024, yao2025roadmap, li_30_2018, rahimi-eichi_battery_2013, tarascon_key_2010, guo_degradation_2021} 
The widespread use of LIBs in consumer electronics and electric vehicles highlights their central role in modern energy storage, where improvements in both electrodes and current collector durability can significantly enhance performance, lifespan, and safety while reducing resource consumption and environmental impact. In the past decade, advances in the exfoliation of layered materials have stimulated significant interest in the incorporation of two-dimensional materials into novel battery designs. \cite{nicolosi_liquid_2013, zhang_2d_2016}. The exceptional electrical conductivity and mechanical strength of MXene layers make them ideal conductive scaffolds and binder materials for incorporating low-dimensional components in both anodes and cathodes.\cite{zhang_interconnected_2023, ahmed_atomic_2017, wang_electrochemical_2021, li_3d_2020}. One such material, SnSe, has attracted attention for use in LIB anodes due to the high theoretical capacity of tin \cite{li_tin_2015, yuan_surfactantfree_2017, wu_sn-based_2021}. Unlike most commercial electrode materials, which operate via lithium intercalation into layered structures,\cite{quilty_electron_2023,li_degradation_2020,li_intercalation_2019, aurbach_common_1998} SnSe undergoes conversion and alloying reactions with lithium to form Li$_{4.4}$Sn, Sn, and Li$_2$Se, enabling a higher lithium storage capacity.\cite{shen_two-dimensional_2023}. However, the associated alloying reactions induce substantial volume expansion, which leads to mechanical instability and poor cyclability, ultimately limiting the practical benefits of the high specific capacity \cite{kim_atomicscale_2018, zhao_yolkshell_2019, zhou_encapsulating_2016}. Incorporating SnSe nanoparticles into MXene sheets offers a promising strategy to address these issues. The mechanical stiffness of the MXene framework can accommodate the large volume changes of SnSe during cycling, reducing structural damage, while its excellent electrical conductivity maintains efficient electron transport throughout the composite.\cite{arunasalam_low-dimensional_2024, zhang_enhancing_2022}.

Achieving long term battery performance requires not only the development of advanced electrode materials but also a deep understanding of how they interact with the current collector during operation. Electrode degradation can result from a complex combination of mechanical, chemical, and electrochemical processes, including structural changes within the active material, modifications of the solid electrolyte interphase (SEI), and breakdown of the current collector interface. Copper, the standard current collector material used for anodes, plays a crucial role in maintaining electrical contact throughout cycling. However, it is increasingly recognized that copper itself can degrade under electrochemical conditions. Copper corrosion and ion migration can lead to electrical disconnection, internal short circuits, and contamination of the electrode and electrolyte.\cite{zhu2021review, guo_degradation_2021} While earlier studies have provided indirect evidence for copper dissolution and redistribution during cycling, direct nanoscale observation of this process has been lacking. Understanding copper corrosion is therefore essential for improving the reliability of both conventional and next generation electrode systems, as current collector breakdown can trigger cascading degradation even when the active material is well designed.

Comprehensive characterization techniques are required to capture structural, compositional, and chemical information while preserving the electrode as close as possible to its operating state.\cite{vetter_ageing_2005, etacheri_challenges_2011}.  Many advanced materials, including those for lithium and sodium ion systems, are highly sensitive to ambient conditions, and the volatility of liquid electrolytes presents additional challenges for high vacuum techniques such as electron microscopy.

Atom probe tomography (APT) has emerged as a powerful tool for investigating battery materials,\cite{kim_atom_2022, li_atomic-scale_2022, yadav2025advancing, singh2023near} including MXenes,\cite{kramer_nearatomicscale_2024} owing to its sub nanometre spatial resolution and high chemical sensitivity. Cryogenic APT (cryo APT) preserves volatile phases such as electrolytes and SEI by maintaining cryogenic conditions during specimen preparation, transfer, and analysis. This enables direct investigation of frozen liquid electrolytes and SEI layers that would otherwise be lost during conventional sample preparation. The combination of cryogenic focused ion beam (cryo FIB) and cryo APT techniques provides the ability to study buried interfaces and degradation mechanisms with unprecedented detail.\cite{schwarz_field_2020, el-zoka_enabling_2020, mccarroll_new_2020, meng_frozen_2022, mulcahy2025workflow} The first demonstration of cryo-APT applied to battery materials was reported by Kim et al. in 2022. \cite{kim_understanding_2022} The development of cryo-FIB lift-out techniques has enabled the preparation of such battery materials containing frozen electrolytes and SEI layers. This approach was initially adapted from workflows established for biological TEM sample preparation. \cite{parmenter_cryofibliftout_2021, rubino_site-specific_2012, parmenter_cryogenic_2014, mahamid_focused_2015}. 

In this study, we investigate the degradation mechanisms of SnSe nanoparticle anodes embedded within a MXene superstructure that we have shown in our previous work as excellent anode material for high-capacity Li-ion batteries.\cite{arunasalam2025high} We initially invesitage the material with cryo FIB slice and view imaging provides a broad view of morphological evolution across the electrode after cycling, revealing pronounced changes such as interfacial roughening, pore formation, and redistribution of material. These observations highlight the complex, depth dependent nature of degradation within the electrode. Guided by these observations, cryo APT was applied to selected regions at various depths in both cycled and uncycled electrodes, enabling atomic scale compositional analysis while preserving volatile and beam sensitive phases. This approach allows direct examination of SEI chemistry, electrolyte infiltration, SnSe dissolution, and, critically, copper corrosion and ion migration from the current collector into the electrode. Together, cryo FIB and cryo APT provide a comprehensive, multiscale framework for linking structural and chemical evolution in battery electrodes. The results not only reveal the nanoscale origins of performance decay in SnSe MXene systems but also present the first direct evidence of copper current collector degradation, demonstrating that even with advanced electrode design, current collector breakdown remains a central factor governing the longevity and safety of lithium ion batteries.

\section{Results and discussion}

\begin{figure}[h!]
    \centering
    \includegraphics[width=\linewidth]{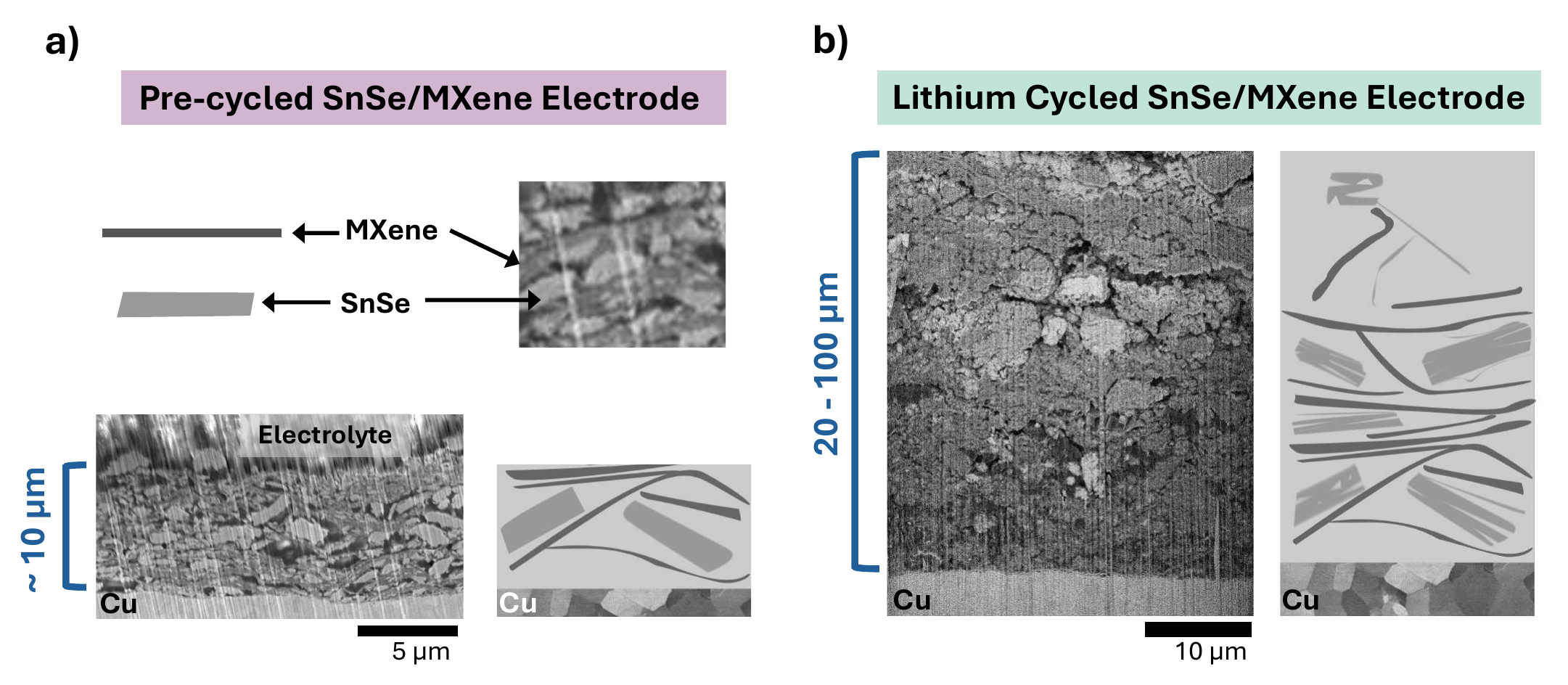}
    \caption{Cryogenic FIB and SEM slice and view imaging of the expansion and damage occurring within the SnSe electrode during cycling showing. (a) SEM image and corresponding schematic of the native SnSe MXene electrode pre-cycling soaked with electrolyte. (b) SEM imaging showing the expansion of the electrode material after lithium electrochmical cycling. All SEM images were taken using the TLD detector in SE-mode}
    \label{fig:Figure_1}
\end{figure}

Due to SnSe alloying with Li during the charge process, significant mechanical damage occurs to the electrode during the cycling process. Cryo FIB SEM slice and view was employed to analyse the SnSe/MXene electrode and electrolyte morphology in three dimensions. This technique combines serial ion beam milling and SEM imaging under cryogenic conditions, enabling depth-resolved visualization of the electrode’s internal structure while preserving the native morphology of beam and temperature sensitive phases such as the electrolyte and SEI. A schematic of an uncycled electrode based on experimental secondary electron SEM imaging  is shown in figure \ref{fig:Figure_1}a. The electrode is composed of nanoparticles of $SnSe$, wrapped in long, thin sheets of $Ti_3C_2$, the MXene binder keeping the material together. The thickness of uncycled electrodes was chosen through the manufacturing process as \SI{10}{\micro\metre}. This is confirmed by the corresponding SEM image of an uncycled electrode, Figure \ref{fig:Figure_1}b, which was soaked in the electrolyte and subsequently frozen. The particles of $SnSe$ are imaged in lighter gray contrast, surrounded by MXene sheets that bind the material together, as indicated by arrows.

Damage to the electrode can arise from multiple, interconnected mechanisms during electrochemical cycling. First, repeated lithiation and delithiation induce substantial volume expansion and contraction within the active material. These cyclic volume changes generate significant mechanical stresses throughout the electrode. To relieve this stress, the SnSe particles crack and fracture into smaller fragments, which in turn increases porosity and leads to partial separation of particles within the composite framework. As a result, the electrode progressively thickens and loses structural integrity. This behaviour is illustrated schematically in Figure~\ref{fig:Figure_1}c.

A secondary electron SEM image from a cryo-FIB slice-and-view series of the bulk electrode after 30 cycles (Figure~\ref{fig:Figure_1}d) shows a thickness of approximately \SI{40}{\micro\metre}, representing a fourfold increase compared to the uncycled state. Across all samples, expansion factors between two- and tenfold were observed via slice and view. 

\begin{figure}[h!]
    \centering
    \includegraphics[width=\linewidth]{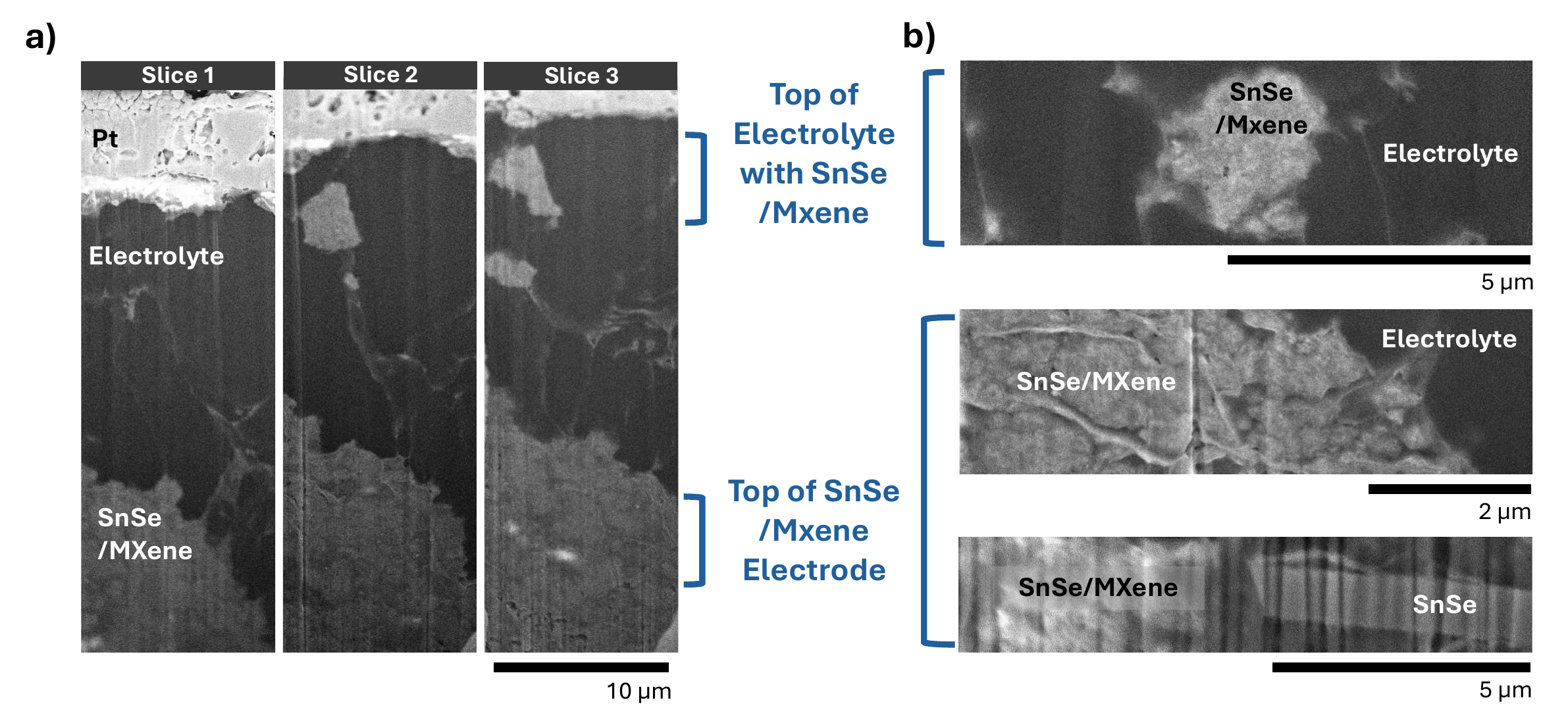}
    \caption{SEM images of cycled SnSe electrodes permeated with electrolyte, showing a) several low magnification slices with a thin capping layer of Pt, significant amounts of electrolyte, and particles of SnSe both at the bottom as well as spread throughout the electrolyte, b) a high magnification image of a mostly intact SnSe particle in the process of delaminating (circled in red), and c) a high magnification image of a mix of SnSe and MXene in electrolyte}
    \label{fig:Figure_2}
\end{figure}

 During cycling, electrodes undergo both significant physical and chemical damage. Figure \ref{fig:Figure_2} a) shows the extent of this damage through several cryogenic FIB slice and SEM view imaging of the electrode. The majority of the SnSe located at the bottom of the slices is covered in a thick layer of electrolyte, and finally a thin layer of Pt deposition from the FIB. However, the electrolyte also shows thin lines of lighter contrast from delaminated material, as well as individual particles located significantly above the main mass of the electrode, and essentially suspended in the frozen electrolyte. Some of these particles are suspended more than \SI{10}{\micro\metre} above the rest of the electrode, larger than the thickness that the dry electrode was originally cast to. These are structures that are not observed in electrodes before cycling. Thus, it is likely that these are seen due to the extreme mechanical stresses of the cycling process, which are significant enough to dislodge large chunks of material from the electrode, which then float around within the electrolyte during the continued cycling process.

While some contrast variation remains, the distinction between individual MXene layers and SnSe particles becomes much less pronounced after cycling compared to the pre-cycled sample. Similarly, the particle edges no longer display sharp contrast at the electrolyte interface but instead exhibit a gradually fading boundary, as seen in Figure~\ref{fig:Figure_2}a and the higher-magnification image in Figure~\ref{fig:Figure_2}b. This behaviour indicates extensive fraying of the particles during cycling, likely caused by a combination of mechanical degradation and partial chemical dissolution into the electrolyte. The high magnification image in Figure~\ref{fig:Figure_2}b shows a largely intact rectangular SnSe particle within the cycled electrode, from which thin flakes of material are detaching a clear indication of the delamination process that occurs during cycling. Both MXene and SnSe are layered materials, and thus delamination can take place in each. Such layered materials lack three-dimensional atomic bonding; instead, they are characterised by strong in-plane covalent interactions and relatively weak van der Waals forces between layers, allowing exfoliation into ultrathin,\cite{nicolosi_liquid_2013, conroy2017importance} high-aspect-ratio nanosheets with very large surface areas. The numerous thin strands observed in Figure~\ref{fig:Figure_2} further demonstrate the material’s high susceptibility to delamination, which is expected to significantly affect the structural stability of the electrode.

\begin{figure}[!h]
    \centering
    \includegraphics[width=0.8\linewidth]{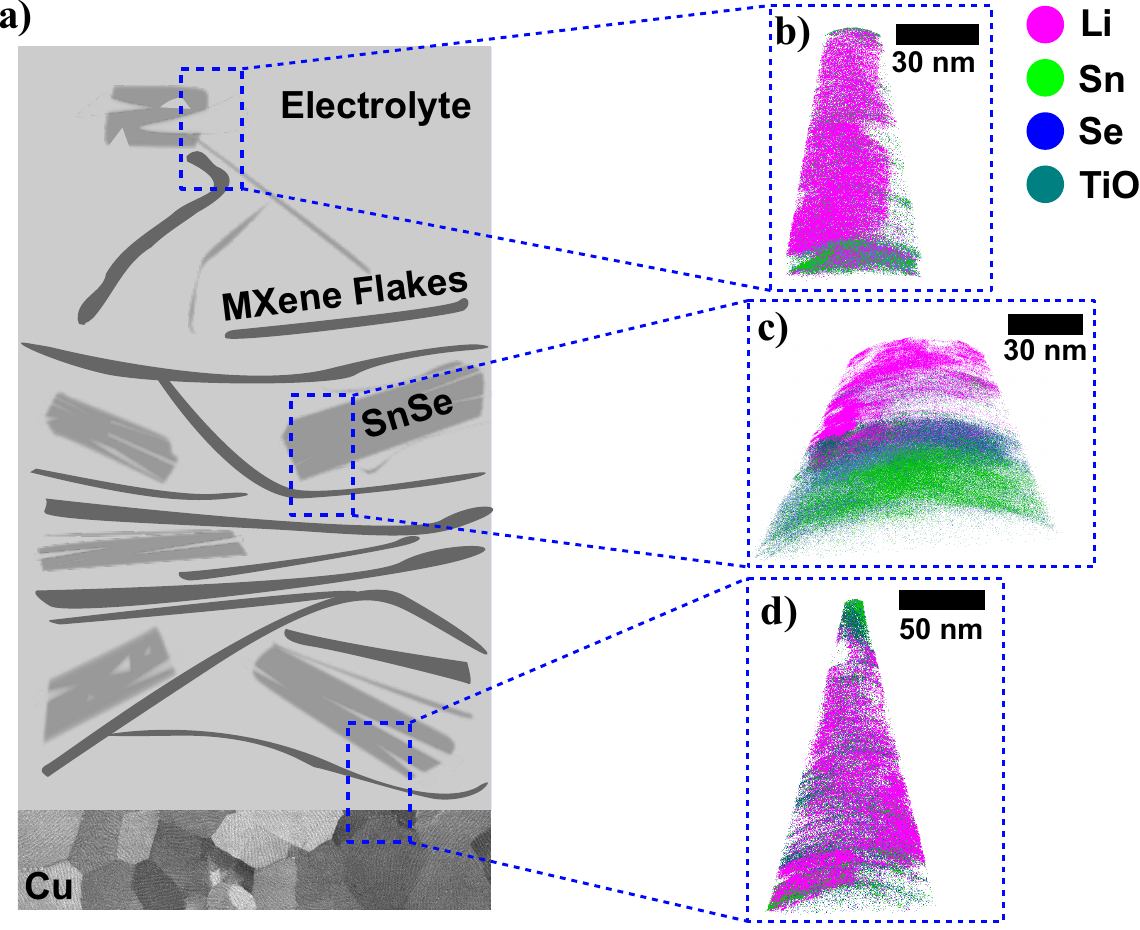}
    \caption{ (a) Schematic of a cycled electrode, showing the principle of taking samples at different depths and reconstructing needles from them, (b) a reconstructed needle from near the top of an electrode, (c) a reconstructed needle from the bulk of an electrode, and (d) a reconstructed needle from the electrode interface with the current collector}
    \label{fig:Figure_3}
\end{figure}

For a more holistic assessment of the damage mechanisms in an electrode, a full picture of processes throughout the entire electrode is required. Specifically, processes near the current collector may differ from those at the top of the electrode. As one can see from \ref{fig:Figure_2}a, some of the electrode material in fact becomes completely detached after cycling and moves multiple microns away into the lithium electrolyte, while some material seems still fully intact such as seen for SnSe particle in \ref{fig:Figure_2}b.  However, the electrode thickness of \SI{10}{\micro\metre}, which significantly exceeds the field-of-view of a typical APT analysis, i.e. within the range of 100 x 100 x 100s of \SI{}{\nano\metre^3}. After cycling, the problem is significantly exacerbated, with a single specimen now only capturing a volume hardly representative of the highly heterogeneous electrode.

Nonetheless, APT still allows for the examination of different parts of the sample through adjusting the liftout and sample preparation process. The use of a plasma FIB allows for the removal of large volumes of material, allowing to expose regions deep below the surface of even thick electrodes during the specimen preparationprocess. Through careful milling, any part of the resulting liftout can be processed into a needled-shaped specimen suitable for subsequent APT analysis. Figure \ref{fig:Figure_3} shows a schematic of a cycled electrode, along with reconstructed atom probe needles that were taken at various depths of sample material. As can be seen, changes in both composition and segregation of the electrode materials can be tracked throughout the depth of the electrode. Quantitative comparison between datasets requires significant care, as the amount of electrolyte in each individual specimen can vary significantly, but different quantitative effects and distributions of material can be tracked.

\begin{figure}[!h]
    \centering
    \includegraphics[width=\linewidth]{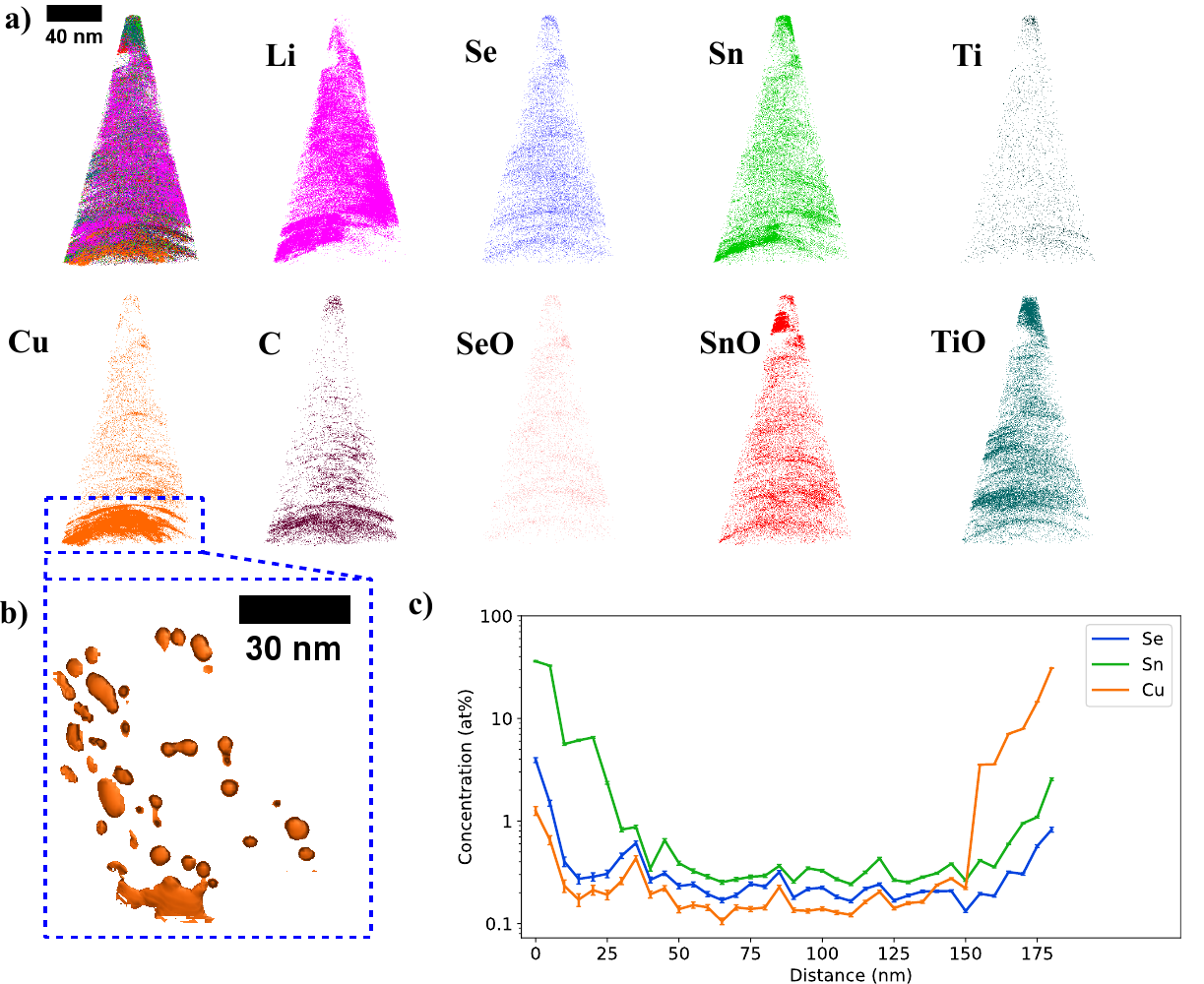}
    \caption{APT Measurement from a needle at the interface with the current collector showing a) the reconstructed needle and individual reconstructions of several different ion species, b) clusters of Cu at the bottom of the specimen from the area marked in blue, rotated to a top-down view, shown as an isosurface at 19 at\%, and c)  concentration profiles of a selection of elements along the z axis of the needle, in log scale.}
    \label{fig:12472Images}
\end{figure}

The needle made from the bottom of the bulk sample, (area from Figure \ref{fig:Figure_3}d) and Figure \ref{fig:12472Images} has a clear measurable Cu-rich signal, indicating that the interface of electrode to current collector has been captured in the measurement. The interface shows micro-scale damage to the current collector in the form of small clusters of Cu-dense regions at the interface. The Cu rich region is shown in Figure \ref{fig:12472Images}b), the increase in Cu at the base of the needle is plotted in \ref{fig:12472Images}c, and a tomography rotational perspective of where the Cu rich region is located is also shown in \ref{fig:CuRotation}. 

While there is a significant amount of Cu from the needle made the current collector interface, that the majority of the needle is composed of Li. The Li which tends to migrate under the strong electrostatic fields used to enable field evaporation during APT measurements \cite{greiwe_atom_2014}, and thus is preferentially pulled from the surface of the tip. The remaining part of the data is composed largely of Sn, Se, and Ti, and C from the electrode materials, as well as Cu from the current collector. Oxide and hydride ions of these are also present, and likely originate during electrode preparation. The absence of nitride species indicates that the cryogenic preparation method successfully prevented nitrogen incorporation from ambient exposure prior to freezing on the cold block. Consequently, the sample remains undamaged by environmental contact, and the measurements reflect the properties of the native material. The organic solvents present in the electrolyte contribute a complex set of peaks, arising from species with diverse compositions of carbon, hydrogen, and oxygen, as detailed in \ref{fig:12287MassSpectrum} and \ref{fig:12472MassSpectrum}.

From the APT analysis the remainder of the needle content is made up of the SnSe and MXene components. The levels of Sn and Se are very similar throughout the needle, as shown in figure \ref{fig:12472Images}c), with the exception of the top of the needle, where there is a region significantly richer in Sn. In general, the Sn in the sample shows more clustering behaviour, whereas the Se is spread more evenly throughout the needle, as shown in \ref{fig:12472NN}. This suggests that after the cycling process, the dissolved SnSe molecules are not able to fully reform, and are partially dissolved into the electrolyte.  There is also measurable increase in SnO and TiO at the top of the needle in Figure \ref{fig:12472Images} with visible clustering throughout the needle.

\begin{figure}[!h]
    \centering
    \includegraphics[width=\linewidth]{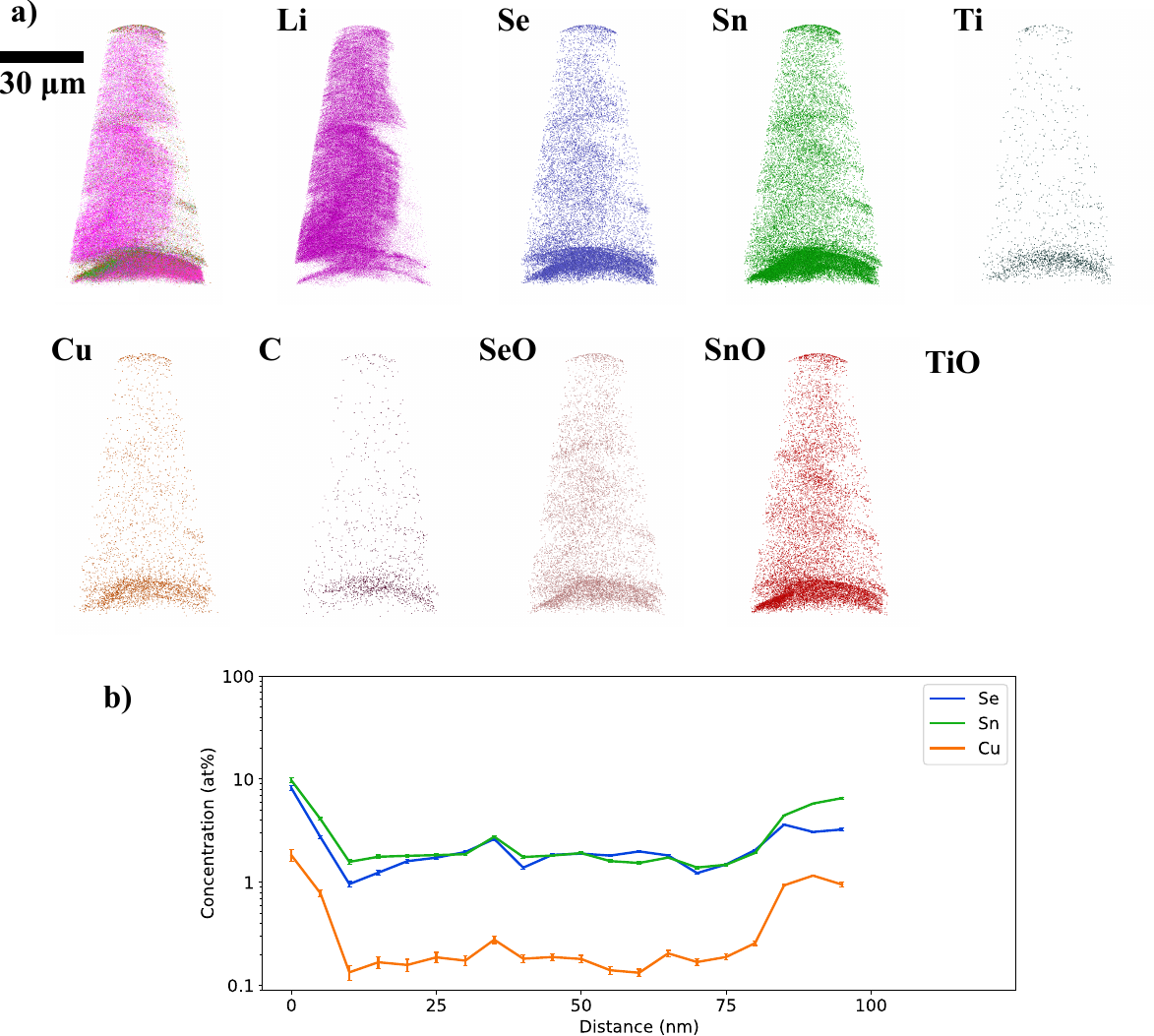}
    \caption{APT measurement from a needle taken from the top of an electrode, showing (a) the reconstructed needle and individual reconstructions of several
different ion species and (b) concentration profile of a selection of elements along the z axis of the needle in log scale.}
    \label{fig:12287Images}
\end{figure}

Figure~\ref{fig:12287Images} presents a reconstruction of a needle extracted from the region near the top of the electrode and electrolyte interface, as illustrated schematically in Figure~\ref{fig:Figure_3}a. As shown previously in Figure~\ref{fig:Figure_1}, the upper regions of the electrode generally exhibit a lower material density, particularly after cycling, which reduces the average amount of electrode material present in each needle. The reconstructed specimen displays markedly less clustering of electrode components, especially tin (Sn). However, a region of solid material remains at the base of the needle, where the clustering of Sn and selenium (Se) is more coherent and better aligned than in the needle taken from the current collector region.

The atomic percentage of Cu increases only in regions where Sn and Se are also present, indicating that Cu enrichment occurs together with these elements and not independently. In contrast to the needle at the current collector interface, the Cu increase in this sample is proportional to the local Sn and Se concentrations, and no isolated Cu-rich regions are observed. The levels of Ti and TiO are much lower in this needle, which is likely due to its greater distance from the nearest part of the MXene support structure. Overall, the improved spatial correlation between Sn and Se suggests that the cycling process affects the regions of the electrode closest to the current collector more strongly. However, the presence of Sn and Se throughout the electrolyte shows that dissolution of the electrode material continues across the full depth of the electrode.

\begin{figure}[!h]
    \centering
    \includegraphics[width=0.90\linewidth]{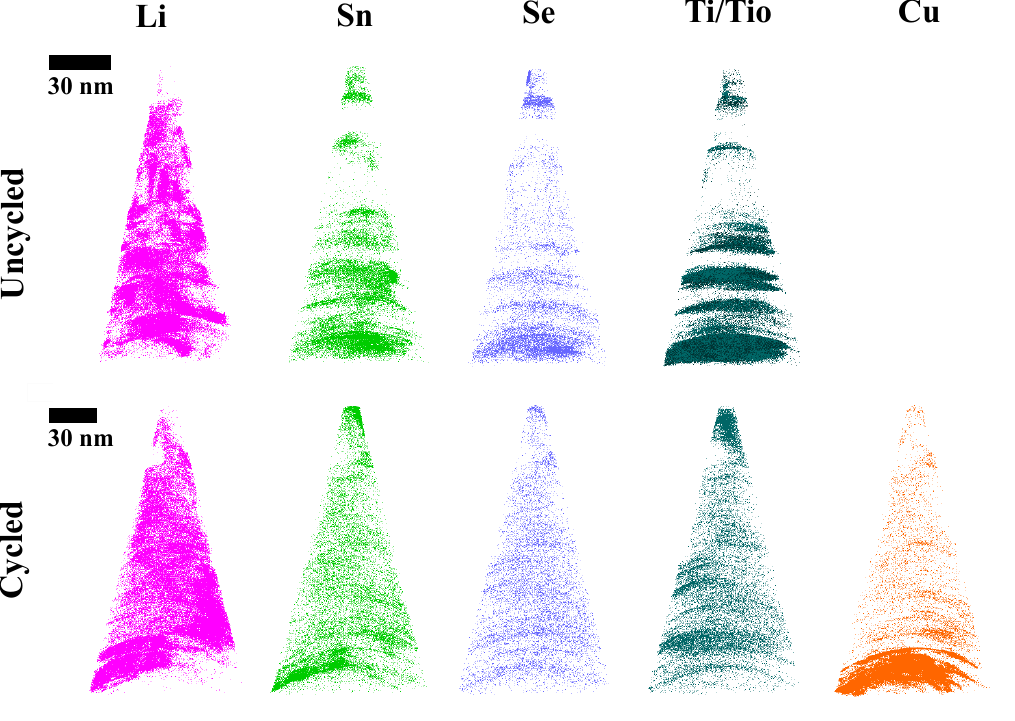}
    \caption{APT measurements of two different needles, one from an uncycled electrode with drop-cast electrolyte and one from a cycled electrode}
    \label{fig:AtomProbeComparison}
\end{figure}

Figure \ref{fig:AtomProbeComparison} shows a comparison of two needles, one made from an uncycled electrode, and one from an electrode extracted from the same cycled cell. In order to allow for construction of the needle and ensure similar baselines between the two needles, the uncycled electrode was dropcast with electrolyte prior to freezing. The uncycled specimens consists of visible layers of electrode material, shown in the maps of Sn, Se, and Ti, interspersed with regions of electrolyte, consisting almost entirely of Li. In comparison, the cycled needle still has regions of increased density of non-Li materials, but there are no clear layers visible. The concentration of Sn, Se, and Ti is much more homogenous in this cycled needle. This shows the direct consequence of the cycling of the electrode, in which Sn and Se alloy with Li, and are then unable to fully recombine to solid particles as the cell is cycled, leading to loss of capacity. 

The Cu signal is either completely absent in the uncycled needles or so weak that it cannot be distinguished from the overlapping TiO peaks~\cite{worch2025fillinggapatomprobe}. In contrast, \textbf{all cycled needles regardless of their distance from the current collector exhibit a clear Cu signal}, providing the \textbf{first direct evidence of copper corrosion and subsequent ion migration} within the system. While previous studies have provided only \textit{indirect evidence} of copper dissolution and redistribution during electrochemical cycling, this phenomenon has remained inferred rather than directly observed~\cite{zhou2025corrosion, guo_degradation_2021, hanf2020accessing, jin2023corrosion, zhu2021review}. The present results therefore establish, for the first time, that cycling-induced damage to the current collector releases Cu ions, which are distributed throughout the electrolyte and can be detected across the full depth of the sample.

\section{Conclusion}
In this work, cryogenic focused ion beam (cryo FIB) slice and view imaging was combined with depth resolved cryogenic atom probe tomography (cryo APT) to investigate the degradation mechanisms in SnSe MXene composite electrodes for lithium ion batteries. Cryo FIB slice and view provided a wide field, three dimensional perspective of morphological evolution during cycling, capturing electrode expansion, pore formation, and redistribution of material across the electrode thickness. These observations revealed that degradation processes vary significantly with depth, driven by local differences in lithium insertion, mechanical stress, and electrolyte infiltration.

Depth resolved cryo APT measurements offered atomic scale insight into these degradation phenomena. The analyses showed incomplete recombination of Sn and Se after cycling, with Sn exhibiting clustering behaviour while Se remained more uniformly distributed, consistent with partial dissolution of SnSe into the electrolyte and limited recovery of the pristine structure. Crucially, a distinct Cu signal was detected throughout all cycled electrodes, even in regions far from the current collector, representing the first direct evidence of copper corrosion and ion migration within this system. This finding provides new insight into the role of current collector degradation as a hidden but significant source of contamination and performance loss in composite electrodes. Electrolyte derived species were also observed throughout the electrode, underscoring the extensive penetration of electrolyte components and their contribution to ongoing interfacial reactions. Together, these results reveal a complex interplay of mechanical, electrochemical, and chemical degradation processes acting over multiple length scales during cycling.

The integration of cryo FIB slice and view with depth resolved cryo APT establishes a powerful multiscale framework for correlating mesoscale structural evolution with nanoscale chemical transformations at buried interfaces. Beyond elucidating the degradation mechanisms of SnSe MXene electrodes, this approach uncovers the previously unobserved copper corrosion pathway, offering a new perspective on current collector stability and its broader implications for the longevity and safety of next generation energy storage materials.

\section{Materials \& Methods}


\subsection{Fabrication of Coin Cells}
SnSe-MXene electrodes were fabricated from SnSe nanoparticles and MXene nanosheets. The synthesis of component materials, fabrication of the composite electrodes, and fabrication of coin cells is described in detail in \citeauthor{arunasalam2025high}\cite{arunasalam2025high}.

\subsection{Electrochemical Testing}

Coin cells were cycled at voltages between \SI{10}{\milli\volt} and \SI{2}{\volt}. Details on the cycling protocol and the resulting curves can be found in \citeauthor{arunasalam2025high}\cite{arunasalam2025high}.


\subsection{Coin Cell Freezing \& Transfer}

Coin cells were opened in an Ar glovebox (UNIlab pro eco, MBraun, Germany) using a coin cell crimper with opening die. A block of brass was cooled just outside the antechamber by partially submerging it in liquid nitrogen, leaving the top of the block exposed. Boil-off from the liquid nitrogen creates a nitrogen atmosphere covering the top of the block . The coin cells were transferred out of the glovebox and immediately placed on the cold block, allowing for a gentle cooling process under moisture-free atmosphere. Once frozen solid, the samples are fully submerged in LN2 and transferred into a nitrogen atmosphere glovebox (Sylatech, Germany) with integrated LN2. From the glovebox, samples are loaded into transfer suitcases (UHVCTS, Ferrovac, Switzerland), capable of transferring samples to other instruments at cryogenic temperatures and vacuum conditions. Schematics of this process can be found in figure \ref{fig:FreezingProcess}.

\subsection{APT Sample Preparation}

Preparation of APT samples was done in a Helios 5 Hydra Dualbeam with integrated cryo-stage (Thermo Fisher Scientific, USA). Samples can be transferred in cryogenically using the mentioned transfer suitcases. Lamellae were lifted out and attached to Si microarrays (Cameca, USA) using in-situ sputtering from the manipulator \cite{douglas_situ_2023} and SEMGlu \cite{mulcahy_look_2025}. Schematics of this can be found in the SI figure \ref{fig:APTPrepProcess}. Sharpening was done primarily at \qty{0}{\degree} tilt with angled mills from the side of the sample, as shown in figure \ref{fig:APTPrepProcess}g. As the porosity of the sample means needles become unstable, even when filled with electrolyte, redeposition of Cr onto one side of the needle was used to strengthen the material during sharpening. This is shown in figure \ref{fig:APTPrepProcess}h.

\subsection{APT}

APT analysis was done in a LEAP 5000 XR (Cameca, USA), with the \qty{355}{nm} laser set to energies between \qtyrange{10}{30}{pJ}. Pulse rate was set to adjust automatically by the system to allow for registration of peaks up to \qty{200}{Da} in order to detect the heavy Sn and Se compounds. Temperatures were kept between \qtyrange{40}{50}{K} to allow for lower voltages, as the samples are very susceptible to fracture.

Reconstructions were made using IVAS 6.3. The migration of Li under the field of the LEAP significantly decreases the spatial accuracy of reconstructions, especially when based on voltage as migration causes voltage to jump frequently. As such, reconstructions were made using fixed shank reconstruction methods. Analysis was carried out with the significantly reduced spatial accuracy in mind. Mass to charge ratio spectra used for the reconstructions can be found in figures \ref{fig:12472MassSpectrum} and \ref{fig:12287MassSpectrum}.

\begin{acknowledgement}

This work was made possible by the EPSRC Cryo-Enabled Multi-microscopy for Nanoscale Analysis in the Engineering and Physical Sciences EP/V007661/1. L.W., K.A. and G.T. acknowledged the EPSRC Centre for Doctoral Training in the Advanced Characterisation of Materials (CDTACM)(EP/S023259/1) for their PhD studentship funding. G.T. acknowledges Cameca Ltd. For co-funding their PhD. SRJ thanks PhD funding from the Faraday Institution, under the grant EP/S514901/1. N.M. and M.S.C. acknowledge funding from Engineering and Physical Sciences Research Council (EPSRC) and Shell for funding through the InFUSE Prosperity Partnership (EP/V038044/1).  M.S.C. acknowledges funding from Royal Society Tata University Research Fellowship (URF\textbackslash R1\textbackslash 201318), Royal Society Enhancement Award\\ (RF\textbackslash ERE\textbackslash 210200EM1) and ERC CoG DISCO grant 101171966. B.G.
acknowledges financial support from the ERC-CoG-SHINE-771602. We thank Teng Zhang and Yury Gogotsi from Drexel University for their Mxene samples.

\end{acknowledgement}
\clearpage

\begin{suppinfo}

\renewcommand{\thefigure}{S\arabic{figure}}
\setcounter{figure}{0}

\begin{figure}[h]
    \centering
    \includegraphics[width=0.6\linewidth]{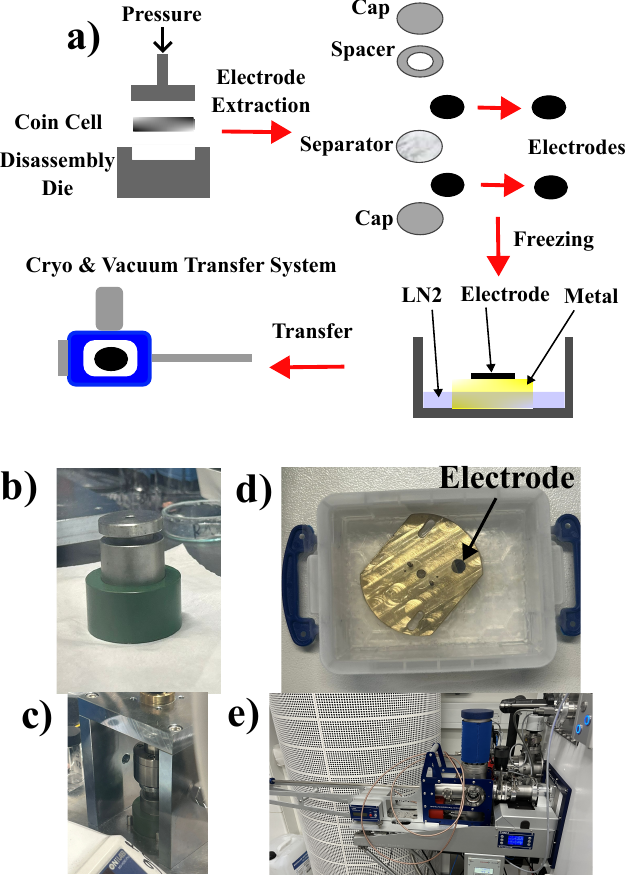}
    \caption{Process of opening coin cells and freezing the electrodes, showing (a) a schematic of the entire process, (b) the decrimping die used to open the cells, (c) the decrimping die within the coin cell crimper, (d) an electrode being frozen on a brass block, partially submerged in liquid nitrogen, (e) the transfer suitcase used to transport samples between instruments}
    \label{fig:FreezingProcess}
\end{figure}

\begin{figure}[p]
    \centering
    \includegraphics[width=\linewidth]{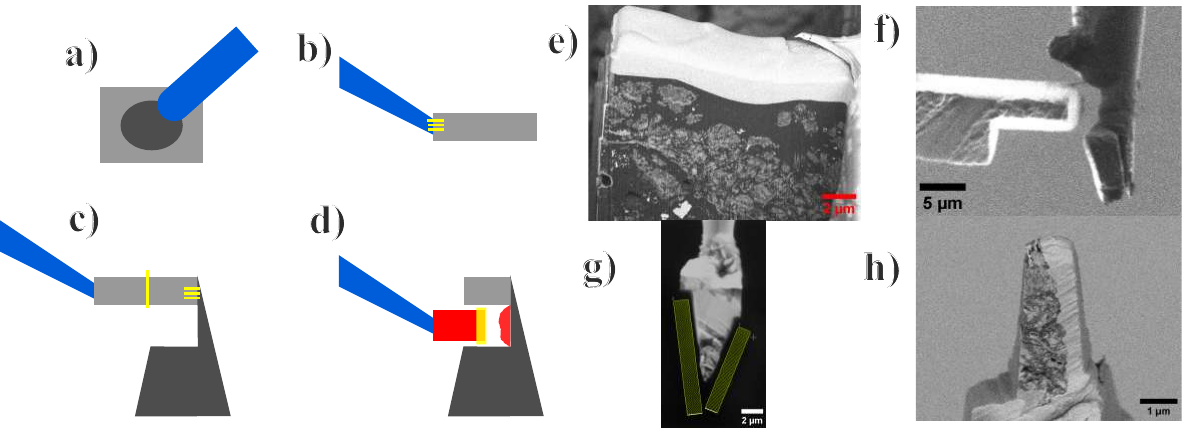}
    \caption{Preparation of APT needles in the SEM-FIB under cryogenic conditions, with milling patterns shown in yellow. Showing (a) the use of the GIS to deposit a precursor gas onto the sample, which is then cured with the electron beam, (b) the use of redeposition welding to attach the micromanipulator to the sample, (c) the use of redeposition welding to attach the sample to an Si post, (d) redepisition of Cr to strengthen the bond of the sample to the Si post, (e) an SEM view of a lamella of SnSe filled with electrolyte, and coated with Pt under cryogenic conditions, (f) and SEM view of the Cr filling procedure, (g) an SEM view of the sharpening of the sample into a needle, and (h) deposition of Cr onto the side of a needle to strengthen the material during sharpening}
    \label{fig:APTPrepProcess}
\end{figure}

\begin{figure}[p]
    \centering
    \includegraphics[width=\linewidth]{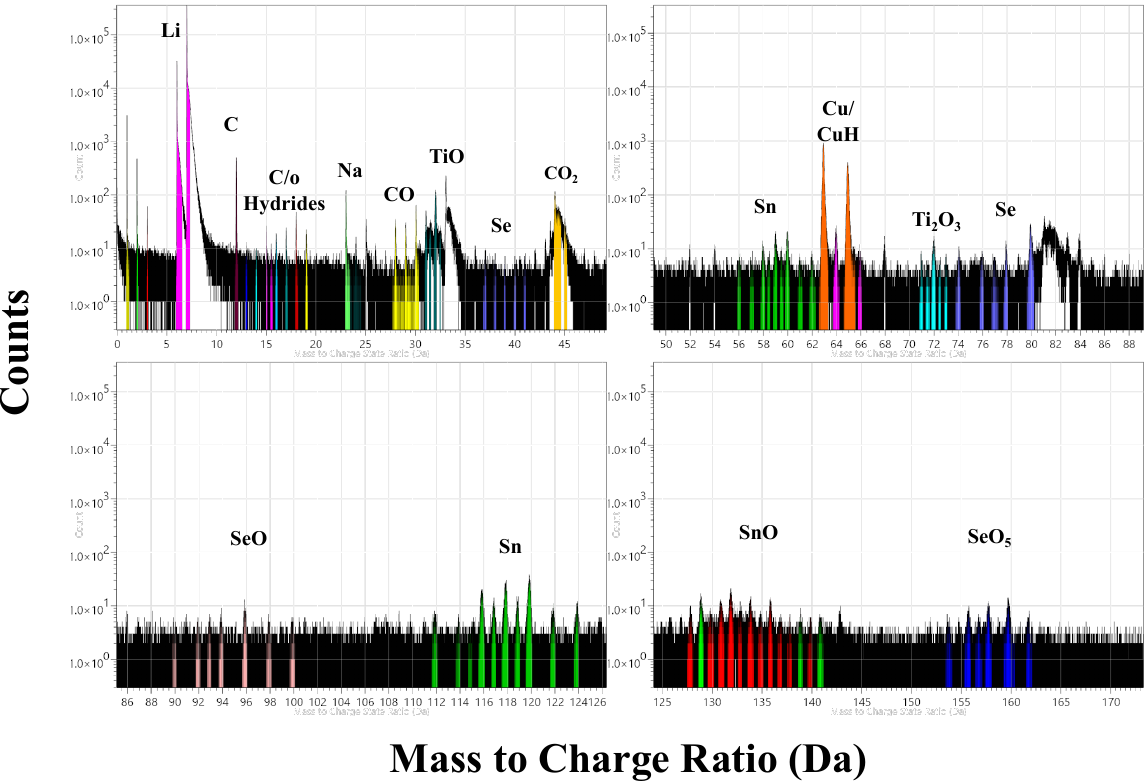}
    \caption{APT mass spectrum of the specimen shown in figure \ref{fig:12472Images}}
    \label{fig:12472MassSpectrum}
\end{figure}

\begin{figure}[p]
    \centering
    \includegraphics[width=\linewidth]{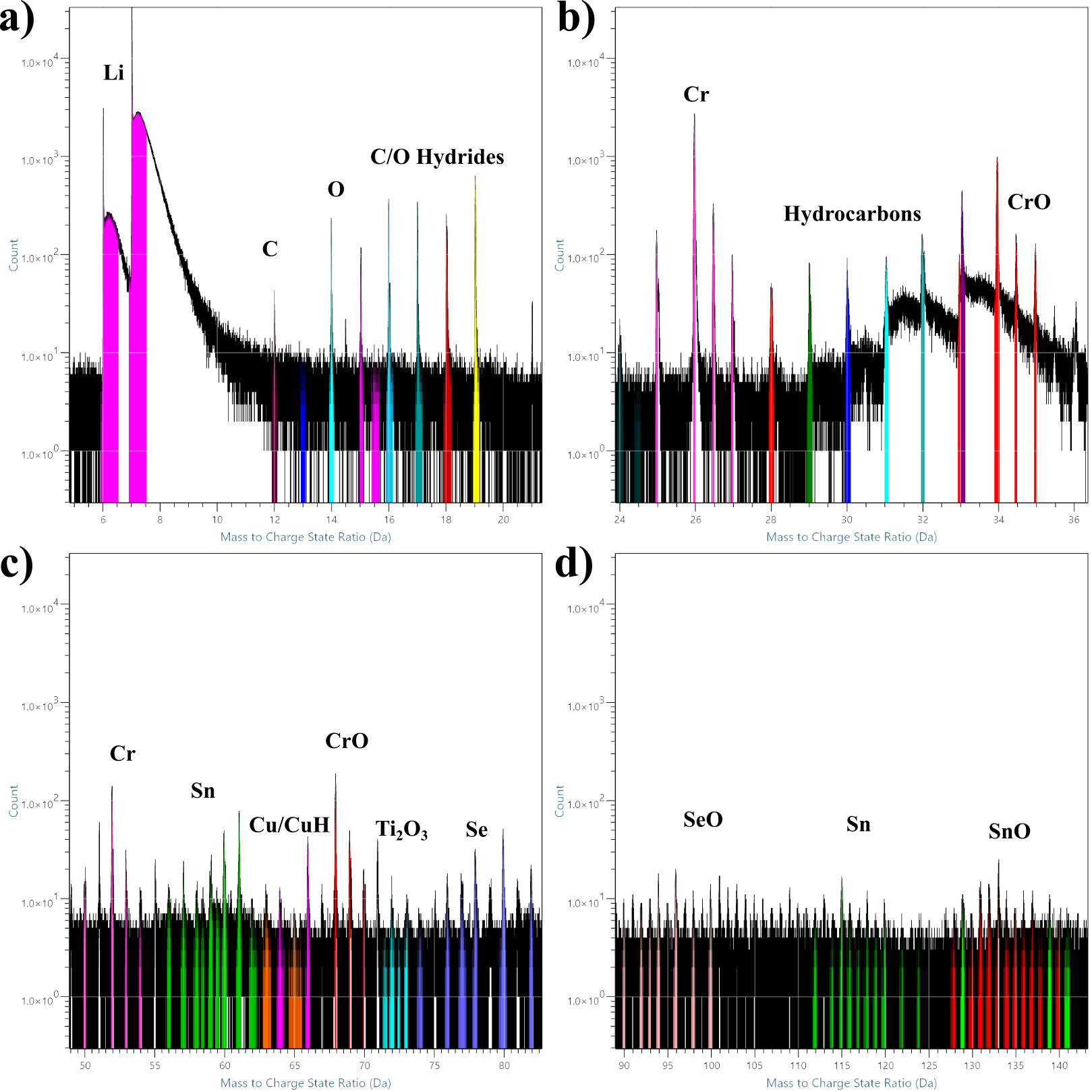}
    \caption{APT mass spectrum of the specimen shown in figure \ref{fig:12287Images}}
    \label{fig:12287MassSpectrum}
\end{figure}

\begin{figure}
    \centering
    \includegraphics[width=\linewidth]{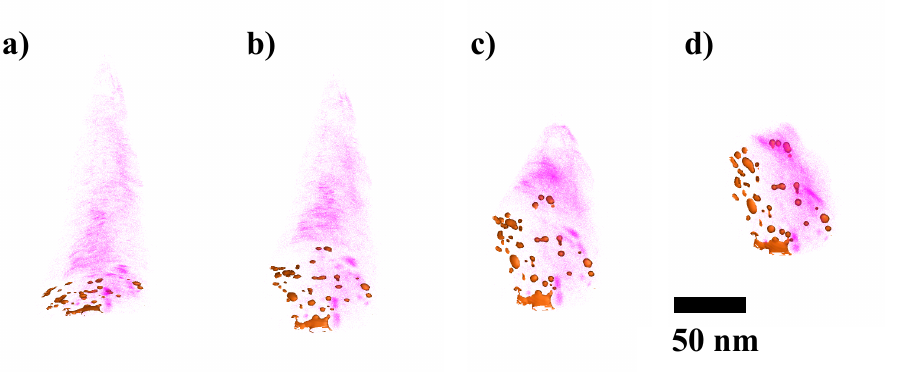}
    \caption{Location of the Cu particles shown in figure \ref{fig:12472Images} in relation to the Li in the needle, shown as a) a sideward view, b) a view rotated \ang{30} toward a top-down view, c) a \ang{60} rotation in the same direction, and d) a top-down view. The Li is purposefully left transparent to show the Cu particles more clearly}
    \label{fig:CuRotation}
\end{figure}

\begin{figure}
    \centering
    \includegraphics[width=\linewidth]{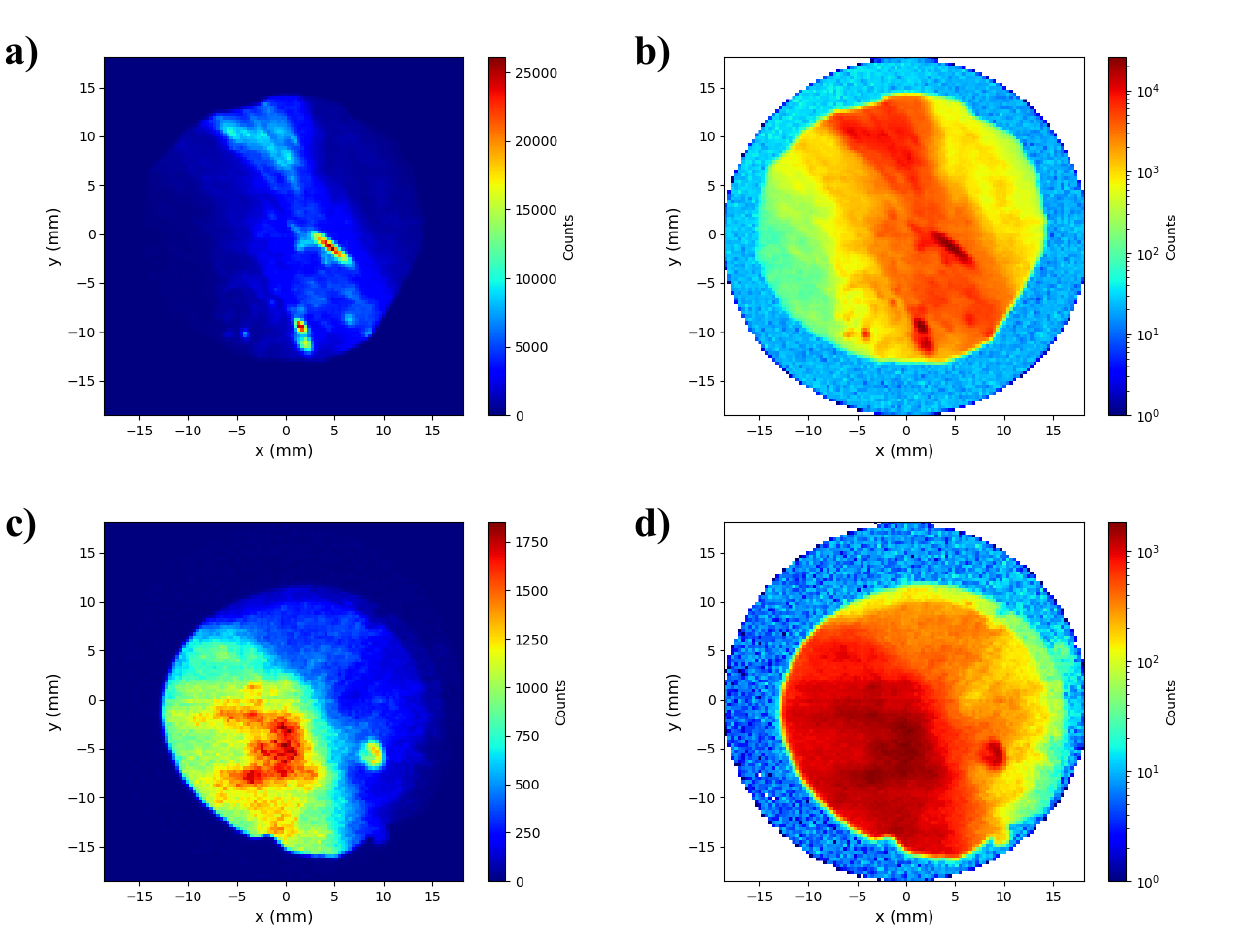}
    \caption{Detector hitmaps showing a) the hitmap of the reconstruction in figure \ref{fig:12472Images}, b) the same hitmap with a log scale, c) the hitmap of the reconstruction in figure \ref{fig:12287Images}, and d) the same hitmap in log scale}
    \label{fig:Detectors}
\end{figure}

\begin{figure}
    \centering
    \includegraphics[width=\linewidth]{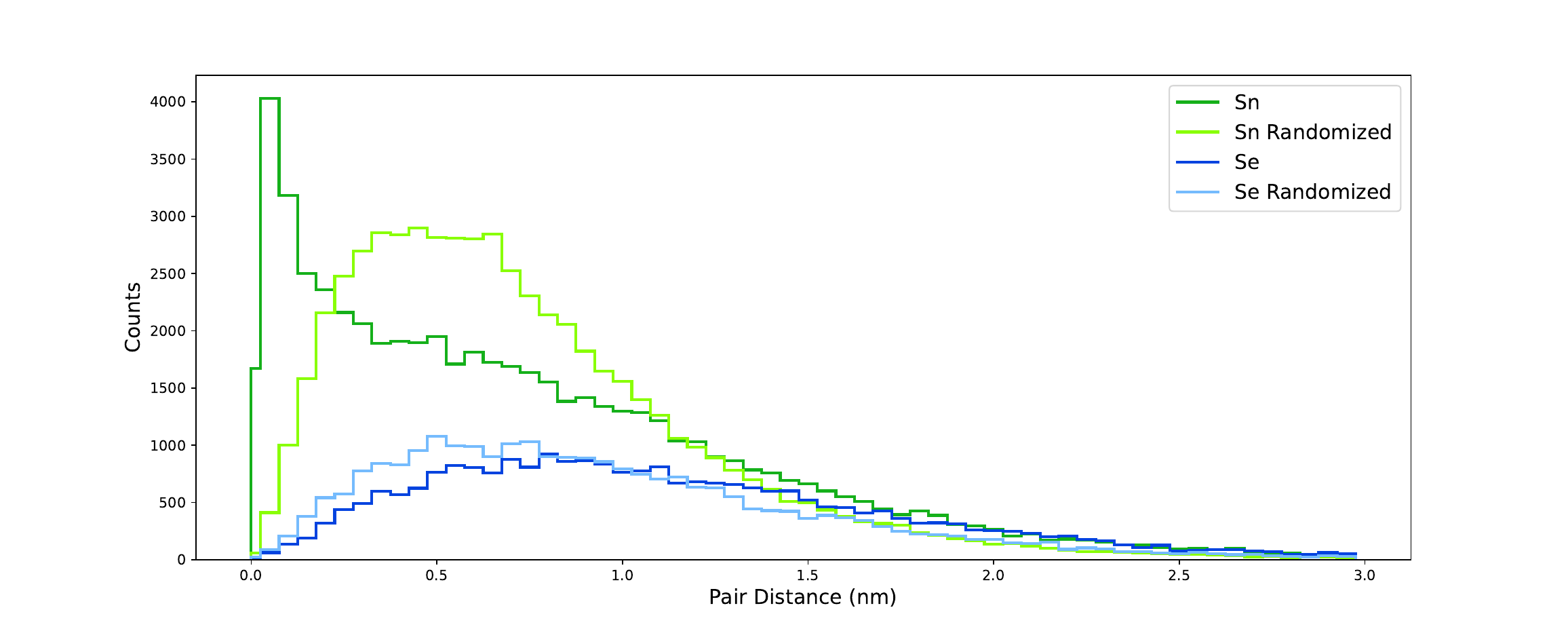}
    \caption{Nearest neighbour distribution of the Sn and Se containing species found in figure \ref{fig:12472Images}, showing both the actual distributions, as well as expected distributions if the same number of ions were randomly distributed. Clustering behaviour in the Sn species can be very clearly seen, whereas the Se distribution is much closer to the randomized state.}
    \label{fig:12472NN}
\end{figure}
\clearpage
\begin{figure}
    \centering
    \includegraphics[width=0.9\linewidth]{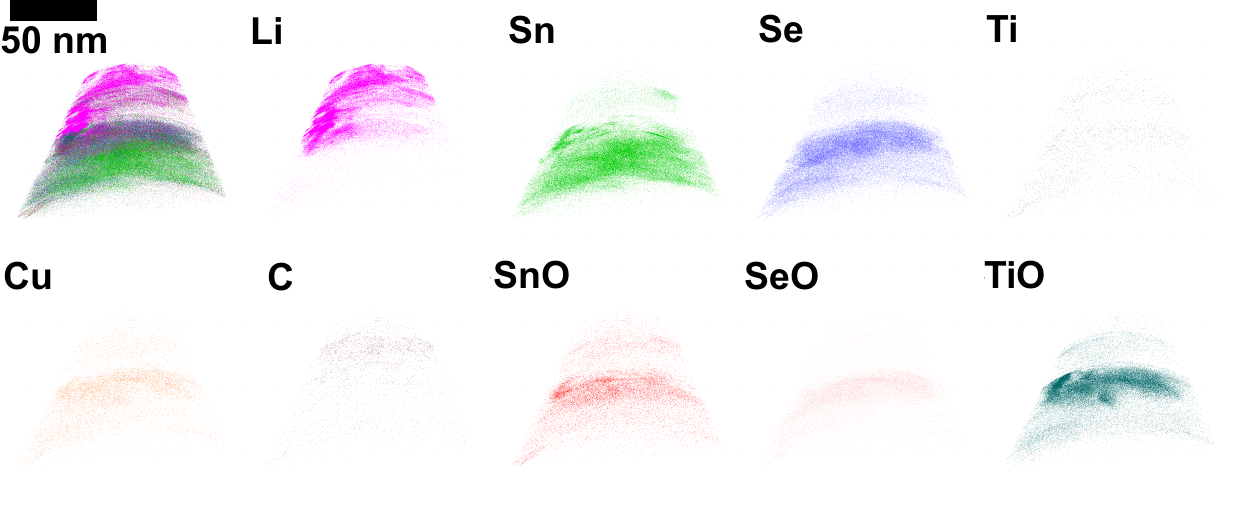}
    \caption{APT measurement from a needle near taken from the middle of an electrode, showing reconstructions of the individual ion species}
    \label{fig:placeholder}
\end{figure}

\end{suppinfo}
\clearpage

\bibliography{bib.bib}

\end{document}